\documentclass[10pt,english]{article}
\usepackage[T1]{fontenc}
\usepackage[latin9]{inputenc}
\usepackage{amsmath}
\usepackage{amssymb}
\usepackage{graphicx}
\usepackage{esint}
\usepackage{feyn}

\makeatletter

\providecommand{\tabularnewline}{\\}

\usepackage{amsfonts} 
\newcommand{\eq}{\begin{equation}}
\newcommand{\eqn}[1]{\label{#1}\end{equation}}
\newcommand{\eea}{\end{eqnarray}}
\newcommand{\eqa}{\begin{eqnarray}}
\newcommand{\eqan}[1]{\label{#1}\end{eqnarray}}
\newcommand{\ba}{\begin{array}}
\newcommand{\ea}{\end{array}}
\newcommand{\eqac}{\begin{equation}\begin{array}{rcl}}
\newcommand{\eqacn}[1]{\end{array}\label{#1}\end{equation}}

\usepackage{babel}

\makeatother

\usepackage{babel}
\begin{document}

\title{\textbf{Remarks on Gribov mechanism on N=1 Supersymmetric 3D
theories and the possibility of obtaining Gribov from one ABJM like Theory.}\\
 \textbf{ }}

\author{\textbf{M.~M.~Amaral}%
\thanks{mramaciel@gmail.com%
}\,\,, \textbf{V.~E.~R.~Lemes}%
\thanks{vitor@dft.if.uerj.br%
}\,\,\\[2mm] \textit{\small{{{{{{UERJ $-$ Universidade do
Estado do Rio de Janeiro}}}}}}}{\small{{{{{}}}}}}\\
 {\small{{{{{ {} }}}}}}\textit{\small{{{{{{Instituto
de Fsica $-$ Departamento de Fsica Terica}}}}}}}{\small{{{{{}}}}}}\\
 {\small{{{{{ {} }}}}}}\textit{\small{{{{{{Rua S{}o
Francisco Xavier 524, 20550-013 Maracan{}, Rio de Janeiro, RJ,
Brasil.}}}}}}}}

\maketitle
\vspace{-1cm}

\begin{abstract}

Some remarks on Gribov mechanism on N=1 Supersymmetric 3D
theories are presented. The two point correlation function is analysed and the possibility of obtaining the confining Gribov regime is discussed. Also the possibility of obtaining Gribov behaviour in ABJM is presented.

\end{abstract}
\setcounter{page}{0}\thispagestyle{empty}

\vfill{}
 \newpage{}\ \makeatother

\section{Introduction}

Three-dimensional Yang-Mills (YM) theory is one important model in which it is possible to
investigate nonperturbative aspects of gauge field theories such as color confinement.
The theory has local degrees of freedom and the coupling constant is
dimensionful. This properties indicates that this theory can be seen as an approximation for the high temperature phase of QCD
with the mass gap in the role of the magnetic mass. It is important to note that in $D=3$ it is always possible to introduce a Chern-Simons term \cite{Deser:1982vy, Deser:1981wh}. This term provides a topological mass, opening the possibility for a deconfining phase due to the appearance of a massive excitation not present in the dual superconductor picture \cite{Polyakov:1976fu,Cornwall:1999xw,Cornwall:2001ni}.
Color confinement is a challenging issue that has been studied in different approaches. 
One of them comes from the analysis of copies of Gribov \cite{Gribov}, known generally as 
Gribov problem\footnote{see \cite{Sobreiro:2005ec} for a pedagogical review}, 
which highlights the Gribov-Zwanziger model (GZ) \cite{Zwanziger1,Zwanziger2,Zwanziger25,Zwanziger3}  and its refined version (RGZ) \cite{rgzmodel}. 
One of the Gribov mechanism properties is that it generates propagators for gauge fields with complex poles being impossible 
their identification with the propagation of particle physics, which is interpreted as confinement 
and known generally as Gribov-Zwanziger scenario.
As is widely known the presence of the Gribov problem is a general characteristic of the quantization procedure of Yang-Mills theories. In this procedure occur the existence of Gribov copies that are, in fact, a general property of all the local covariant renormalizable gauge fixing \cite{singer}. The presence of gauge copies results in zero modes of the Faddeev-Popov operator that makes the usual Faddeev-Popov construction incomplete.
So the Gribov mechanism is an interesting possibility  of the investigation of confinement in YM theories in three dimensions.
That is, YM-CS theory with the Gribov mechanism has the possibility of a confined and a de-confined phase, depending on a fine tune between the Gribov mass gap and the usual parameters in the pure YM-CS \cite{SorellaYMCSGribov}. In this case we have an interesting model with different phases.

In this context supersymmetry has been used as an important tool in the investigation of nonperturbative aspects of gauge theories. In confining behavior of supersymmetric theories, $N = 1$ for example, much has been done since the work of Seiberg with super QCD. See reference \cite{terningmodersusy} and references therein that examine in detail the recent developments.
And we can point out yet that in three dimensions there is currently a renewed interest in supersymmetry because it was recently found the correspondence known as $ADS_{4}/CFT_{3}$ \cite{Maldacena}, whose main example is the ABJM theory \cite{Maldacena}, which is closely related to CS theory in the gauge sector. This correspondence is the realization that certain theories of supergravity in higher dimensions are dual of supersymmetric quantum field theory in lower dimensions, and thus makes it possible to relate the strong coupling region of supersymmetric quantum field theory with weak coupling of the supergravity.

It is also important to note that due to the supersymmetric extension is also possible to introduce fermions in a natural way. Of course this fermions are the gauge partners and are not in the fundamental representation as quarks. In spite of that they can be useful in order to understand the behaviour of confining fermions and the possibility of  relations under Gribov and supersymmetric theories.

So in this paper we investigate the Super-Yang-Mills Chern-Simons (SYM-CS) theories ($N = 1, D = 3$) with superfields
formalism \cite{siegelsuperspace} addressing the Gribov problem and the Gribov two-point correlator
\begin{equation}
\mathcal{G}(p^{2})=\frac{p^{2}}{p^{4}+\gamma^{4}}\label{eq:propagatorlikeGribov}
\end{equation}
as well as a modified Gribov Zwanziger type correlator 
\begin{equation}
\mathcal{G}(p^{2})=\frac{p^{2}+M^{2}}{p^{4}+(M^{2}+m^{2})p^{2}+(M^{2}m^{2}+\gamma^{4})},\label{eq:propagatorlikeRGZ}
\end{equation}
and thus obtaining information on how
the theory (SYM-CS) behaves in the presence of Gribov horizon and how to obtain the Gribov regime in a closest relation to the ABJM scenario.

The paper is organized as follows: in Section 2, the SYM-CS theory in superspace $D=3$, $N=1$ is presented, Landau gauge fixing is performed and the Gribov problem is analysed. In section 3 the Gribov Zwanziger local action is presented and a mathematical analysis of the value of the Gribov parameter is presented. In section 4 a possible  mechanism in order to obtain a Gribov behaviour in a ABJM type theory is presented.

\section{Superfield approach to Gribov problem, N = 1, D = 3, SYM-CS theory.}

\subsection{N = 1, D = 3, Euclidean SYM-CS theory}

In three-dimensional Minkowski space-time the Lorentz group is $SL(2,R)$
(instead of $SL(2,C)$) and the corresponding fundamental representation
acts on a two components real spinor (Majorana) . So to formulate
the superspace, we started with the introduction of spinorial coordinates
$\theta^{\alpha}$ (with $\alpha=1,2$) that are transformed under
$SO(1,2)$. In the case of Euclidean $D = 3$, the two components spinor
 shall be transformed under $SO(3)$ and as is well known 
\cite{Kugo:1982bn,McKeon:2000qm,McKeon:2001pm} one can not have the usual
Majorana condition. It's the same question we are in D = 4 \cite{Amaral:2013uya}.
In the same way we follow the approach of generalizing the concept
of complex conjugation of Grassmann algebra \cite{Wetterich:2010ni}. 
The notations and conventions are in Appendix A.

In $D = 3$ it is possible to add an additional gauge invariant term
beyond the YM, the term CS \cite{Deser:1982vy,Deser:1981wh},
which is a topological mass term for the gauge field. Thus the pure
supersymmetry $N = 1$ version of this action must have both terms. Let
us take the Euclidean version of this superspace action of SYM-CS \cite{RuizRuiz1}:
\begin{equation}
S_{SYMCS}=S_{SYM}+S_{SCS},\label{eq:actionSYMCS}
\end{equation}
with,
\begin{equation}
S_{SYM}=\frac{1}{2}\int d^{3}xd^{2}\theta W^{a\alpha}W_{\alpha}^{a},\label{eq:actionSYM}
\end{equation}
and
\begin{equation}
S_{SCS}=im\int d^{3}xd^{2}\theta\left[(D^{\alpha}\Gamma^{a\beta})(D_{\beta}\Gamma_{\alpha}^{a})+\frac{2}{3}igf^{abc}\Gamma^{a\alpha}\Gamma^{b\beta}(D_{\beta}\Gamma_{\alpha}^{c})-\frac{1}{6}g^{2}f^{abc}f^{cde}\Gamma^{a\alpha}\Gamma^{b\beta}\Gamma_{\alpha}^{d}\Gamma_{\beta}^{e}\right].\label{eq:actionSCS}
\end{equation}
The field strength is given by:
\begin{equation}
W_{\alpha}^{a}=D^{\beta}D_{\alpha}\Gamma_{\beta}^{a}+igf^{abc}\Gamma^{b\beta}D_{\beta}\Gamma_{\alpha}^{c}-\frac{1}{3}g^{2}f^{abc}f^{cde}\Gamma^{b\beta}\Gamma_{\beta}^{d}\Gamma_{\alpha}^{e},
\end{equation}
and superspace derivative:
\begin{equation}
D_{\alpha}=\frac{\partial}{\partial\theta^{\alpha}}+i\sigma_{\alpha}^{\mu\gamma}\varepsilon_{\gamma\beta}\theta^{\beta}\partial_{\mu}.
\end{equation}

The supermultiplet of gauge fields is given by the components of the
spinor superfield, in Wess-Zumino gauge:
\begin{align}
\Gamma_{\alpha}^{a}(x,\theta) & =i\sigma_{\alpha}^{\mu\gamma}\varepsilon_{\gamma\beta}\theta^{\beta}A_{\mu}^{a}(x)+i\theta^{2}\lambda_{\alpha}^{a}(x).\label{eq:spinorsuperfieldcomponent}
\end{align}
They belong to the adjoint representation of the gauge group SU(N). 

The classical action for SYM-CS theory, $S_{SYMCS}$, remains invariant
under the following gauge transformation 
\begin{equation}
\delta_{\Lambda}\Gamma_{\alpha}^{a}=(\nabla_{\alpha}\Lambda)^{a},\label{eq:gaugetransformation}
\end{equation}
with superspace covariant derivative: 
\begin{equation}
\nabla_{\alpha}^{ab}=\delta^{ab}D_{\alpha}+gf^{acb}\Gamma_{\alpha}^{c}.\label{superspacecovariantderivative}
\end{equation}

\subsection{Gauge-fixing}

In order to quantize the theory correctly we have to fix the gauge
and we can do covariantly using the usual procedure of Faddeev-Popov
(FP) \cite{siegelsuperspace,RuizRuiz1,West}.

In the supersymmetric Landau gauge we must implement the conditions
$D^{\alpha}\Gamma_{\alpha}^{a}=0$. 
And following the usual procedure we ended with the action of gauge
fixing
\begin{equation}
S_{gf}=\frac{1}{4}s\{\int d^{3}xd^{2}\theta(c'^{a}D^{\alpha}\Gamma_{\alpha}^{a})\},\label{eq:actiongaugefix}
\end{equation}
where the Faddeev-Popov ghost fields will be scalar superfield .
$c'^{a}$ and $c^{a}$ are the antighost and the ghost respectively.
And $s$ is the BRST nilpotent operator ($s^{2}=0)$.

The total action $S=S_{SYMCS}+S_{gf}$ is invariant under the BRST
transformations \cite{RuizRuiz1}:
\begin{align}
s\Gamma_{\alpha}^{a} & =(\nabla_{\alpha}c)^{a}\nonumber \\
sc^{a} & =-\frac{i}{2}gf_{abc}c^{b}c^{c}\nonumber \\
sc'^{a} & =b^{a}\nonumber \\
sb^{a} & =0,\label{eq:brst}
\end{align}
with $s$ carrying ghost number 1.

The ghost part of gauge fixing action becomes:
\begin{equation}
S_{FP}=-\frac{1}{4}\int d^{3}xd^{2}\theta(c'D^{\alpha}\nabla_{\alpha}^{ab}c^{b}),\label{eq:actionghost}
\end{equation}
with superspace covariant derivative given by (\ref{superspacecovariantderivative}):

In order to calculate the propagator for the gauge superfield $\Gamma_{\alpha}$,
we need only the bilinear of S. So, for the bilinear part, we have: 
\begin{align*}
S_{SYM2} & =\frac{1}{2}\int d^{3}xd^{2}\theta(D^{\beta}D^{\alpha}\Gamma_{\beta}^{a})(D^{\gamma}D_{\alpha}\Gamma_{\gamma}^{a})\\
 & =\frac{1}{2}\int d^{3}xd^{2}\theta\Gamma_{\beta}^{a}D^{\alpha}D^{\beta}D^{\gamma}D_{\alpha}\Gamma_{\gamma}^{a},
\end{align*}
and using (\ref{eq:epsoncontraction}, \ref{eq:superderivativealfabetarelation}, \ref{eq:realtionderivativecontraction}, \ref{eq:D2squared}):
\begin{align*}
S_{SYM2} & =\int d^{3}xd^{2}\theta\Gamma_{\beta}^{a}D^{2}D^{\gamma}D^{\beta}\Gamma_{\gamma}^{a},
\end{align*}
and for Chern-Simons:
\begin{align*}
S_{SCS2} & =im\int d^{3}xd^{2}\theta\left[(D^{\alpha}\Gamma^{a\beta})(D_{\beta}\Gamma_{\alpha}^{a})\right]\\
 & =im\int d^{3}xd^{2}\theta\left[\Gamma_{\beta}^{a}D^{\gamma}D^{\beta}\Gamma_{\gamma}^{a}\right].
\end{align*}
So, with $A=D^{2}+im$:
\[
(AD^{\gamma}D^{\beta}+\frac{1}{\xi}D^{\beta}D^{\gamma})(a_{1}D_{\beta}D_{\lambda}+a_{2}D_{\lambda}D_{\beta})=\delta_{\lambda}^{\gamma},
\]
where $\xi$ is a gauge parameter to be set to zero after having evaluated
the gauge propagator. Using (\ref{eq:superderivativealfabetarelation},
\ref{eq:D2squared}, \ref{eq:epsoncontraction}, \ref{eq:realtionderivativecontraction})
and (\ref{eq:superderivativealfabetaalfarelationzero}), (in Landau
gauge $\xi\rightarrow0$, $a_{2}=0$) we get the inverse
\[
\frac{1}{2A\partial^{2}}D_{\beta}D_{\lambda},
\]
so the massive gauge propagator for SYM-CS:
\begin{equation}
<\Gamma_{\alpha}^{a}(1)\Gamma_{\beta}^{b}(2)>=\frac{\delta^{ab}}{\partial^{2}(-\partial^{2}+m^{2})}(D^{2}-im)D_{\beta}D_{\alpha}\delta^{2}(\theta_{1}-\theta_{2})\delta^{3}(x_{1}-x_{2}).\label{eq:SYMCSpropagator}
\end{equation}

\subsection{Gribov problem}

In the previous section we showed generally the quantization of SYM-CS theory by Faddeev-Popov method. 
But one should be careful with this procedure. In the usual YM theory, 
although the gauge be fixed by the Faddeev-Popov method, Gribov showed
in \cite{Gribov} that there are still field configurations obeying the Landau gauge
linked by gauge transformations, \textit{i.e.} there are still equivalent
configurations, or copies, being taken into account into the Feynman
path integral. In other words, the gauge is not completely fixed and
the remaining ambiguity is allowed due to the existence of normalizable
zero-modes of the Faddeev-Popov operator, 
\begin{equation}
\mathcal{M}^{ab}=-\partial_{\mu}D_{\mu}^{ab}\,.
\end{equation}
So we should investigate how this problem appears in the SYM-CS case and its consequences. 
Before we give a brief review of the YM case in d dimensions.

\subsubsection{The Gribov problem in YM theories in d dimensions}

To address the problem of Gribov copies (eliminate these copies)
Gribov showed that the domain of integration
of the functional integral should be restricted to a certain region
$\Omega$, the so-called Gribov region, that is defined as the set
of field configurations performing the Landau gauge condition, for
which the Faddeev-Popov operator is strictly positive, namely 
\begin{equation}
\Omega:=\{\, A_{\mu}^{a}\,|\,\partial_{\mu}A_{\mu}^{a}=0,\,\mathcal{M}^{ab}(A)>0\,\}\,.\label{definicao}
\end{equation}
Its boundary, $\partial\Omega$, where the first vanishing eigenvalue
of the Faddeev-Popov operator shows up, is known as the Gribov horizon.\\

 As in the region $\Omega$ the Faddeev-Popov operator is positive
than its inverse must diverge when approaching the horizon, due to
the existence of a zero mode. So the restriction to the first Gribov
region is implemented requiring that 
\begin{equation}
G(p^{2},A)=\frac{\delta^{ab}}{N^{2}-1}\langle p|(-\partial_{\mu}D_{\mu}^{ab})^{-1}|p\rangle\,,
\end{equation}
which is the normalized trace of the ghost connected two point function
in momentum space, has no pole for a given non vanishing value of the
momentum $p$, except for the singularity at $p=0$, corresponding
to the first Gribov horizon. At $p\approx0$ one can write 
\begin{eqnarray}
G(p^{2},A) & \approx & \frac{1}{p^{2}}\frac{1}{1-\sigma(p^{2},A)}\,,\label{eq:twopointghosparam}\\
\sigma(p^{2},A) & = & \frac{N}{N^{2}-1}\frac{1}{p^{2}}\int\frac{d^{d}q}{(2\pi)^{d}}\frac{(p-q)_{\mu}p_{\nu}}{(p-q)^{2}}A_{\mu}^{a}(-q)A_{\nu}^{a}(q).\label{defregion}
\end{eqnarray}
From the above expression (\ref{defregion}), it follows that the
no-pole condition at finite non vanishing $p$ is 
\begin{equation}
\sigma(p^{2},A)<1.\label{eq:nopolecondictionGribov}
\end{equation}
As $\sigma(p^{2},A)$ decreases as $p^{2}$ increases one can also
take 
\begin{equation}
\sigma(0,A)=\frac{1}{4}\frac{N}{N^{2}-1}\int\frac{d^{d}q}{(2\pi)^{d}}\frac{1}{q^{2}}(A_{\mu}^{a}(-q)A_{\mu}^{a}(q))\leq1\,.\label{sigma}
\end{equation}
 
In order to perform the restriction to the Gribov region into the
partition function, $\mathcal{Z}$, the final step is to introduce
the no-pole condition with the help of a Heaviside function: 
\begin{equation}
{\cal {Z}}=\int{\cal {D}}A\delta(\partial A)\theta(1-\sigma(0,A))\exp^{-S_{\mathrm{YM}}}.
\end{equation}
where the Euclidean $SU(N)$ Yang-Mills action in the Landau gauge ($S_{\mathrm{YM}}$) in d dimensions is given by: 
\begin{equation}
S_{\mathrm{YM}}=\int d^{d}x\,\left(\frac{1}{4}F_{\mu\nu}^{a}F_{\mu\nu}^{a}+ib^{a}\,\partial_{\mu}A_{\mu}^{a}+\overline{c}^{a}\partial_{\mu}D_{\mu}^{ab}c^{b}\right)\,.\label{YM}
\end{equation}

 Note that the only allowed singularity at (\ref{eq:twopointghosparam})
is at $p^{2}=0$, whose meaning is that of approaching the horizon,
where $G(p^{2},A)$ is singular due to the appearance of zero modes
of the Faddeev-Popov operator. Thus we have to take \cite{Gribov}:
\begin{equation}
\sigma(0,A)=1.\label{eq:sigma=00003D00003D00003D1}
\end{equation}
And thus the Gribov parameter $\gamma$ is fixed by the gap equation, 
which is, in the case of $d = 3$ e.g.,
\begin{equation}
\frac{2Ng^{2}}{3}\int\frac{d^{3}q}{(2\pi)^{3}}\frac{1}{q^{4}+\gamma^{4}}=1.\label{gap}
\end{equation}

It is clear that the Gribov approach is only the first step in order
to consistently treat the problem of zero modes and the Gribov copies
in a gauge fixed Yang-Mills theory. The second step is the GZ theory
\cite{Zwanziger25,Zwanziger3}, which consists in a renormalizable
and local way to implement the restriction to the first Gribov region.
In fact, Zwanziger observed that the restriction could be implemented
by adding the following term in the action (\ref{YM}): 
\begin{equation}
S_{\mathrm{GZ}}=S_{\mathrm{YM}}+\gamma^{4}H(A)\,,
\end{equation}
where, $H(A)$ is the so-called horizon function, 
\begin{equation}
H(A)=g^{2}\int d^{d}x\, d^{d}y\, f^{abc}A_{\mu}^{b}(x)[\mathcal{M}^{-1}]^{ad}(x,y)f^{dec}A_{\mu}^{e}\,
\end{equation}
where ${\cal M}^{-1}$ stands for the inverse of the Faddeev-Popov operator.
In the Zwanziger approach, the parameter $\gamma$ is fixed by the
equation 
\begin{equation}
\langle H(A)\rangle=dV(N^{2}-1)\,,\label{gapGZ}
\end{equation}
where $V$ is the Euclidean space volume. 

It is clear that the horizon function is nonlocal, but it can be localized
with the help of a suitable set of auxiliary fields. In order to ensure
that those extra fields do not introduce extra degrees of freedom
they are introduced in the form of a BRST quartet
\begin{eqnarray}
 &  & s{\bar{\omega}}_{\mu}^{ab}={\bar{\varphi}}_{\mu}^{ab}\,,\qquad s{\bar{\varphi}}_{\mu}^{ab}=0\,,\nonumber \\
 &  & s\varphi_{\mu}^{ab}=\omega_{\mu}^{ab}\,,\qquad s\omega_{\mu}^{ab}=0\,,\label{brsgz}
\end{eqnarray}
where $(\bar{\varphi},\varphi)$ are a pair of complex commutating
fields, while $(\bar{\omega},\omega)$ are anti-commutating ones.
Now, the local version of the GZ action is then given by: 
\begin{eqnarray}
S_{\mathrm{GZ}}^{\mbox{{\it local}}} & = & S_{\mathrm{YM}}+s\int d^{d4}x\,\Bigl[\,\bar{\omega}_{\mu}^{ac}\mathcal{M}^{ab}\varphi_{\mu}^{bc}\Bigl]+\int d^{d}x\,\Bigl[\,\gamma^{2}gf^{abc}A_{\mu}^{a}(\varphi_{\mu}^{bc}-\bar{\varphi}_{\mu}^{bc})+dV(N^{2}-1)\gamma^{4}\Bigl]\,.\label{LocalGZ}
\end{eqnarray}
The last term is a vacuum term  permitted by power counting and  required to obtain the gap equation (\ref{gapGZ}) by the condition that the vacuum energy, $\mathcal{E}$, be independent of $\gamma^{2}$, \textit{i.e.},
\begin{equation}
-\frac{\partial\mathcal{E}}{\partial\gamma^{2}}=0\,,
\end{equation}
where the $-\mathcal{E}$, is defined by
\begin{equation}
e^{-\mathcal{E}}=\int[d\Phi]e^{-S_{GZ}},
\end{equation}
and $[d\Phi]$ stands for the integration over all the fields. 

\subsubsection{SYM-SC and Gribov problem}
 
As discussed in the introduction, this problem of Gribov copies is a general
 property of all the local covariant renormalizable gauge fixing  \cite{singer}.
And we explicitly have shown that Gribov problem also
exists in SYM in Landau gauge \cite{Amaral:2013uya} in $D=4$. 
So let's investigate how this problem appears in the gauge fixing procedure of SYM-CS theory.

First we note that in the Landau-gauge the gauge condition is not ideal. In fact if we consider two equivalents superfield, $\Gamma_{\alpha}^{a}$ and $\Gamma_{\alpha}^{a'}$, connected by a gauge transformation (\ref{eq:gaugetransformation}), if both satisfy the same condition of the Landau gauge, $D^{\alpha}\Gamma_{\alpha}^{a}=0$ and $D^{\alpha}\Gamma_{\alpha}^{a'}=0$, we have
\begin{equation}
D^{\alpha}(\nabla_{\alpha}\Lambda)^{a}=0.\label{FPoperatorSusyini}
\end{equation}
Therefore, the existence of infinitesimal copies, even after FP quantization is related to the presence of the zero modes of  the operator above.
This operator is the same that results in FP  action (\ref{eq:actionghost}) and so, if he has zero mode issue there are problem with functional generator 
\begin{equation}
Z=\int D\Gamma e^{-S_{SYMSC}(\Gamma)}.\label{eq:funcional}
\end{equation}
This suggests that we should further restrict the functional integration to a region free of  zero modes, 
and consecutively free of gauge superfields copies.
To do this we would like to study the operator (\ref{FPoperatorSusyini}) in terms of the eigenvalues and eigenvectors equation, which is not immediate since the equation $D^{\alpha}(\nabla_{\alpha}\Lambda)^{a}=\lambda\Lambda$ is not an eigenvalue equation, which can be seen in components. It relates the component at $\theta=0$ with to component $\theta^{2}$ of superfield $\Lambda$. This indicates that the correct operator, where one can study the  zero modes problem, and thus define the Gribov problem is:
\begin{equation}
{\cal O}^{ab}=D^{2}D^{\alpha}\nabla_{\alpha}^{ab}.\label{eq:fFPoperatorSusy}
\end{equation}
This operator is the correct generalization of the FP operator\footnote{We note that the usual Faddeev-Popov operator is the component $\theta^{2}$
of the supersymmetric FP operator that appears (\ref{eq:actionghost}).}  since the  action (\ref{eq:actionghost}) 
has  an integral in $\theta^{2}$.
So to see the zero mode problem  we take the eigenvalues equation
\begin{equation}
D^{2}D^{\alpha}\nabla_{\alpha}^{ab}\Lambda=\lambda\Lambda
\end{equation}
So for configurations close to the vacuum $\Gamma_{\alpha}=0$ and using (\ref{eq:D2squared})
\begin{equation}
-\partial^{2}\Lambda=\lambda\Lambda,
\end{equation}
which has only positive eigenvalues $\lambda=p^{2}>0$, since the operator in question is Hermitian.
However as we go considering larger amplitudes than the vacuum, ie, $\Gamma_{\alpha}$ sufficiently large, this can not be guaranteed and may be displayed negative eigenvalues.
So we can consider the restriction of functional integration for the region free of zero modes of this operator, which will be the generalization of the Gribov region (\ref{definicao})
\begin{equation}
\Omega:=\{\,\Gamma_{\alpha}^{b}\,|\, D^{\alpha}\Gamma_{\alpha}^{b}=0,\,\mathcal{{\cal O}}^{ab}(\Gamma_{\alpha})>0\,\}\,.\label{susyGribovregion}
\end{equation}

In order to implement this restriction then we consider the GZ approach \cite{Zwanziger1,Zwanziger2,Zwanziger25,Zwanziger3} where is included in the functional integral the inverse of this operator (horizon function) in order to compensate the problem,  this is formally
\begin{align}
H(\Gamma_{\alpha}^{a})=\gamma^{4}\int d^{2}\theta\int d^{3}x\; d^{3}y\; f^{abc}\Gamma_{\alpha}^{b}(x)\left[\frac{\varepsilon^{\alpha\beta}}{D^{2}D^{\alpha}\nabla_{\alpha}}\right]^{ad}(x,y)f^{dec}\Gamma_{\beta}^{e}(y)\;.\label{hf1-1}
\end{align}
In the next section we will put this horizon function in your local form and study some implications showing its consistency. The calculation of the field configurations that corresponds to the solutions of the Gribov condition is an extensive work and requires further studies.

\section{Gribov-Zwanziger local action on superspace}

To localize the horizon function (\ref{hf1-1}) 
we observe that we can get rid of the inverse of the FP operator  
with the aid of standard formula for Gaussian integration 
and the introduction of a pair of spinor superfields of bosonic character and other fermionic, ending with
\begin{align}
S_{aux} & =tr\int d^{3}xd^{2}\theta\left[-(w'_{\gamma}\varepsilon^{\gamma\beta}D^{2}D^{\alpha}\nabla_{\alpha}w_{\beta})+(u'_{\gamma}\varepsilon^{\gamma\beta}D^{2}D^{\alpha}\nabla_{\alpha}u_{\beta})+2\gamma^{2}\Gamma_{\gamma}\varepsilon^{\gamma\beta}(u'_{\beta}-u{}_{\beta})\right],\label{eq:superGZaction-1}
\end{align}
There exists a freedom of redefinition of fields and we can perform a linear shift on the field $w_{\beta}$
in order to write the first two terms as a BRST variation. 
This can be seen best by introducing of the
auxiliary spinor superfields in the form of one quartet of BRST:
\begin{align}
sw'_{\alpha}=u'_{\alpha}, & \; su_{\alpha}=w_{\alpha}\nonumber \\
su'_{\alpha}=0, & \; sw_{\alpha}=0,\label{eq:brst-1}
\end{align}

At this point is important for our construction to show the canonical
dimension and ghost number of all fields and operators which are in
Table \ref{table1}.

\begin{table}[h]
\centering %
\begin{tabular}{|c|c|c|c|c|c|c|c|c|c|c|c|}
\hline 
fields and operators  & $\theta^{\alpha}$  & $D_{\alpha}$  & $\Gamma_{\alpha}$  & $c'$  & $c$  & $b$  & $w'_{\alpha}$  & $w_{\alpha}$  & $u'_{\alpha}$  & $u_{\alpha}$  & $g$ \tabularnewline
\hline 
Canonical dimension  & -$\frac{1}{2}$  & $\frac{1}{2}$  & 0  & $\frac{3}{2}$  & $-\frac{1}{2}$  & $\frac{3}{2}$  & 0  & 0  & 0  & 0  & $\frac{1}{2}$\tabularnewline
\hline 
Ghost number  & 0  & 0  & 0  & -1  & 1  & 0  & -1  & 1  & 0  & 0  & 0\tabularnewline
\hline 
\end{tabular}\caption{Quantum numbers of fields and operators.}

\label{table1} 
\end{table}

Thus, we ended up with the following proposal to
the super GZ action:
\begin{align}
S_{aux} & =tr\int d^{3}xd^{2}\theta\left[s(w'_{\gamma}\varepsilon^{\gamma\beta}D^{2}D^{\alpha}\nabla_{\alpha}u_{\beta})+2\gamma^{2}\Gamma_{\gamma}\varepsilon^{\gamma\beta}(u'_{\beta}-u{}_{\beta})\right],\label{eq:superGZaction}
\end{align}
where $\gamma^{2}$ is a mass parameter, which should be determined
by the theory, shown below, must be nonzero. And also include a vacuum term.
The total action is: 
\begin{equation}
S_{SGZ}=S_{SYMCS}+S_{gf}+S_{aux}.\label{eq:totalaction}
\end{equation}

With GZ action generalization at our disposal we can now calculate
the propagators and ensure they have the expected behavior that occurs
in confining YM theories and analyze other features it adds to SYM.

Before starting the calculation of the super propagator it is important to emphasize here that the $N=1$ super Yang-Mills is a renormalizable action in $D=4$ \cite{Capri:2014jqa} and the Gribov procedure for $N=1$, $D=4$ super Yang-Mills was implemented in terms of component fields in \cite{Capri:2014xea} and directly in superspace in \cite{Amaral:2013uya}. Also is well know that Yang-Mills-Chern-Simons is a renormalizable gauge theory and the BRST breaking in the Gribov procedure is a soft breaking that does not spoil the renormalizability, explicitly proven in $D=4$ \cite{Dudal:2009xh,Sorella:2009vt,Sorella:2010fs}. Due to power counting, the Gribov procedure must also be renormalizable in $D=3$.

\subsection{The super gauge propagator}

First we calculate the propagator for gauge superfield $\Gamma_{\alpha}$.
To calculate the gauge propagator we need only the bilinear of S like
to calculate (\ref{eq:SYMCSpropagator}). Thus, for $S_{SGZ}$, we
have (with \ref{eq:D2squared}):
\begin{align}
S_{SGZ2} & =tr\int d^{3}xd^{2}\theta(-u'_{\gamma}\varepsilon^{\gamma\beta}\partial^{2}u_{\beta}+w'_{\gamma}\varepsilon^{\gamma\beta}\partial^{2}w_{\beta}+2\gamma^{2}\Gamma_{\gamma}\varepsilon^{\gamma\beta}u'_{\beta}-2\gamma^{2}\Gamma_{\gamma}\varepsilon^{\gamma\beta}u{}_{\beta}).\label{eq:suzerGZactionbilinear}
\end{align}
With give one contribution to bilinear term 
\begin{align}
S_{SGZ2} & =tr\int d^{3}xd^{2}\theta\Gamma_{\gamma}\frac{2\gamma^{4}}{\partial^{2}}\varepsilon^{\gamma\beta}\Gamma_{\beta}.\label{eq:supergzaction-1-1-1}
\end{align}
Similar to SYM-CS propagator calculus, with $A=D^{2}+im$ and $B=\frac{2\gamma^{4}}{\partial^{2}}$:
\begin{equation}
(AD^{\gamma}D^{\beta}+\frac{1}{\xi}D^{\beta}D^{\gamma}+B\varepsilon^{\gamma\beta})(a_{1}D_{\beta}D_{\lambda}+a_{2}D_{\lambda}D_{\beta})=\delta_{\lambda}^{\gamma},\label{eq:OpInverter}
\end{equation}
and we get the inverse:
\begin{equation}
\frac{1}{2A\partial^{2}+BD^{2}}D_{\beta}D_{\lambda},\label{eq:Opinvertido}
\end{equation}
or
\begin{equation}
-\frac{1}{2}\left[\frac{(\partial^{4}+\gamma^{4})+im\partial^{2}D^{2}}{(\partial^{4}+\gamma^{4})^{2}-m^{2}(\partial^{2})^{3}}\right]D^{2}D_{\beta}D_{\lambda}.
\end{equation}
So the gauge propagator for SYM-CS-GZ:
\begin{equation}
<\Gamma_{\alpha}^{a}(1)\Gamma_{\beta}^{b}(2)>=\frac{1}{2}\delta^{ab}\left[\frac{(\partial^{4}+\gamma^{4})+im\partial^{2}D^{2}}{-(\partial^{4}+\gamma^{4})^{2}+m^{2}(\partial^{2})^{3}}\right]D^{2}D_{\beta}D_{\alpha}\delta^{2}(\theta_{1}-\theta_{2})\delta^{3}(x_{1}-x_{2}).\label{eq:SYMCSpropagatorGZ}
\end{equation}

To see how the introduction of $S_{SGZ}$ brings light on confinement
of both bosons as fermions and to compare with literature, we shall
observe the propagators in field components.

Taking components from (\ref{eq:spinorsuperfieldcomponent})
we can project the propagator for the gauge field $A_{\mu}$ :
\begin{equation}
<A_{\mu}^{a}(x_{1})A_{\nu}^{b}(x_{2})>=\delta^{ab}\left[\frac{(\partial^{4}+\gamma^{4})(-\partial^{2})}{(\partial^{4}+\gamma^{4})^{2}-m^{2}(\partial^{2})^{3}}\right](\delta_{\mu\nu}-\frac{\partial_{\mu}\partial_{\nu}}{\partial^{2}}-\frac{im\partial^{2}\varepsilon_{\mu\nu\sigma}\partial_{\sigma}}{(\partial^{4}+\gamma^{4})})\delta^{3}(x_{1}-x_{2}),
\end{equation}
and gaugino $\lambda^{\alpha}$:
\begin{equation}
<\lambda_{\alpha}^{a}(x_{1})\lambda_{\beta}^{b}(x_{2})>=\frac{1}{4}\delta^{ab}\left[\frac{(\partial^{4}+\gamma^{4})}{(\partial^{4}+\gamma^{4})^{2}-m^{2}(\partial^{2})^{3}}\right](\partial^{2}\partial_{\beta\alpha}-\frac{im(\partial^{2})^{3}\varepsilon_{\beta\alpha}}{(\partial^{4}+\gamma^{4})})\delta^{3}(x_{1}-x_{2}).
\end{equation}

And we found that both show behavior in limit $m=0$ as occurs for
gauge field in non-supersymmetric theories (\ref{eq:propagatorlikeGribov}).
However as we'll see in the next section the Gribov parameter $\gamma$
can be determined as a function of coupling constant $g$ such that
these propagators will be function of the two parameters, $g$ and
$m$ (in the case $m\neq0$), and so it is possible the study of phases
involving this theory. The phases for this type of propagator was
studied in \cite{SorellaYMCSGribov} through the analysis of the poles in this propagator. It is important to note here that due to a Gribov type propagator also for the fermionic partners we have a fermionic condensate defined by $\lim_{x_{1}\rightarrow x_{2}}Tr <\lambda_{\alpha}^{a}(x_{1})\lambda_{\beta}^{b}(x_{2})>$. This condensate is responsible for maintaining the energy equal to zero and compensate the gauge condensate that usually appears in GZ. In simple terms, the existence of the fermion condensate ensures that the supersymmetry is preserved in spite of breaking the BRST symmetry as usually happens in GZ.

\subsection{Ghost propagators and $\gamma$ parameter}

Since the action (\ref{eq:totalaction}) only makes sense if the $\gamma$
parameter is nonzero, we will now explicitly show that it is not independent
in this theory. Its determination is closely linked to the restriction
of the functional integration to the first Gribov region, which we
will discuss some details here.

First, it is noteworthy that in the literature dealing with the Gribov
problem in YM theories there are recent consensus on the scenario
of dominance of configurations on the Gribov horizon on the Landau
gauge \cite{reviewZwanziger}, so that the restriction to the first
Gribov region is, in practice, to take the configurations on the horizon,
ie where occur the zeros modes of the FP operator. Second, and as
we have pointed out in the introduction of super GZ, calculate the
propagator of the ghosts is to take the inverse of these operators.
So we focus on these calculus to one loop order to establish the one
loop gap equation Gribov style.

In order to characterize the integration in the first Gribov region
it is important to remember that the two point ghost function is essentially
the inverse of the Faddeev-Popov operator and the zero eigenvalue
of the Gribov equation corresponds to a exactly to the Gribov frontier.
In these sense the two point ghost function goes to infinity at the
Gribov frontier. These condition is the most simpler way to obtain
the gap equation for $\gamma$. These procedure is explained in details
in \cite{Gribov} and is easily extended to the $N=1$ supersymmetric
case \cite{Amaral:2013uya}. First we need to calculate the two
point ghost function. Using perturbation theory these is at first
order of the form:

\begin{center}
\begin{tabular}{ccc}
 &  & \tabularnewline
$\Diagram{\\
\vertexlabel_{c'_{a}}h & \momentum{hA}{p} & h\vertexlabel_{c{}_{b}}
}
$  & +  & $\Diagram{ &  & \momentum{glA}{k}\\
\vertexlabel_{c'_{a}}\momentum{hA}{p} & h & \momentum{hA}{p-k} & \momentum{hA}{p}\vertexlabel_{c{}_{b}}
}
$\tabularnewline
 &  & \tabularnewline
\end{tabular}
\par\end{center}

Where the line between $c'_{a}$ and $c_{b}$ corresponds to the zero
order super ghost propagators $\mathcal{G}_{c'c}^{ab0}=-\triangle_{c'c}^{c}(1,2)$.
After a straightforward calculation:
\begin{equation}
\mathcal{G}_{c'c}^{ab0}=\frac{-2}{p^{2}}\delta^{ab}D^{2}\mbox{\ensuremath{\delta^{2}}(\ensuremath{\theta_{1}}-\ensuremath{\theta_{2}})}.\label{eq:ghostpropagator1}
\end{equation}

And we can define in momentum space, the one loop corrected ghost
propagator as

\begin{equation}
\mathcal{G}_{c'c}^{ab}=(\mathcal{G}_{c'c}^{ab0}+\mathcal{G}_{c'c}^{ab1}),\label{eq:ghostpropzeroplusoneorder}
\end{equation}
according to diagram above. With $\mathcal{G}_{c'c}^{ab0}$ given
from (\ref{eq:ghostpropagator1}).

Using the Feynman rules and D algebra from \cite{siegelsuperspace,Srivastava,PetrovQuantumsuperfield},
(and due $f^{acd}f^{bcd}=N\delta^{ab}$):
\begin{multline}
\mathcal{G}_{c'c}^{ab1}=(2\pi)^{3}g^{2}N\delta^{ab}\int d^{2}\theta_{3}\int d^{2}\theta_{4}\frac{1}{p^{2}}D_{1}^{2}\delta^{2}(\theta_{1}-\theta_{3})\\
D_{3}^{\alpha}\left\{ \int\frac{d^{3}k}{(2\pi)^{3}}\frac{(k^{4}+\gamma^{4})}{(k^{4}+\gamma^{4})^{2}+m^{2}k^{6}}\frac{1}{(p-k)^{2}}\left(1+\frac{imk^{2}D_{3}^{2}}{(k^{4}+\gamma^{4})}\right)D_{3}^{2}D_{\beta}D_{\alpha}\delta^{2}(\theta_{3}-\theta_{4})D_{3}^{2}(p-k)\delta^{2}(\theta_{3}-\theta_{4})\right\} \overleftarrow{D_{4}^{\beta}}\\
\frac{1}{p^{2}}D_{4}^{2}\delta^{2}(\theta_{4}-\theta_{2}).
\end{multline}
And after delta functions and D derivatives manipulations \cite{PetrovQuantumsuperfield},
we have:
\begin{align}
\mathcal{G}_{c'c}^{ab1} & =-4(2\pi)^{3}g^{2}N\delta^{ab}\frac{1}{p^{2}}D_{1}^{2}\delta^{2}(\theta_{1}-\theta_{2})\int\frac{d^{3}k}{(2\pi)^{3}}\left(\frac{(k^{4}+\gamma^{4})}{(k^{4}+\gamma^{4})^{2}+m^{2}k^{6}}\right)\frac{k^{2}}{(p-k)^{2}}.
\end{align}
Next we define:
\begin{align}
\sigma(\gamma^{2},p^{2},m^{2}) & =2(2\pi)^{3}g^{2}N\int\frac{d^{3}k}{(2\pi)^{3}}\left(\frac{(k^{4}+\gamma^{4})}{(k^{4}+\gamma^{4})^{2}+m^{2}k^{6}}\right)\frac{k^{2}}{(p-k)^{2}}.
\end{align}
Therefore, from (\ref{eq:ghostpropzeroplusoneorder}):
\begin{align}
\mathcal{G}_{c'c}^{ab} & =-2\delta^{ab}\frac{1}{p^{2}}D_{1}^{2}\delta^{2}(\theta_{1}-\theta_{2})(1+\sigma).
\end{align}
Re-summing the one-particle irreducible diagrams gives:
\begin{align}
\mathcal{G}_{c'c}^{ab} & =-2\delta^{ab}\frac{1}{p^{2}}D_{1}^{2}\delta^{2}(\theta_{1}-\theta_{2})\frac{1}{(1-\sigma)}.
\end{align}

Now, as we are interested in the low momentum behavior we analyze
the behavior of $(1-\sigma)$ we get:
\begin{align}
\sigma(\gamma^{2},0,m^{2}) & =2(2\pi)^{3}g^{2}N\int\frac{d^{3}k}{(2\pi)^{3}}\left(\frac{(k^{4}+\gamma^{4})}{(k^{4}+\gamma^{4})^{2}+m^{2}k^{6}}\right).\label{eq:integralD3equacaogapcomparametrodeCS}
\end{align}

According to the above discussion of the scenario of dominance of
configurations on the Gribov horizon, ie the ghost propagator (the
inverse of FP operator) going to infinity, $(1-\sigma)=0$, we have
to get the greatest value of the above integral witch is with $m=0$
as we can see in the integral graph shown in Figure \ref{figintegral}:
\begin{center}
\begin{figure}[!h]
\center
\caption{Graph of integral $\sigma(\gamma,m)$ (\ref{eq:integralD3equacaogapcomparametrodeCS})}
\includegraphics[scale=0.45]{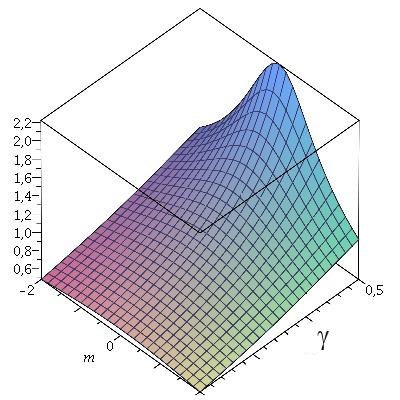}
\label{figintegral}
\end{figure}
\end{center} 
At this point is important to remember that also in the Gribov procedure applied to non supersymmetric Yang-Mills it is necessary to analyse the greatest value of the integral $\sigma$. In the non supersymmetric Yang-Mills in $D=4$ we also take the integral at zero external momenta, which corresponds to the greatest value of the integral $\sigma$. It is easy to note in Figure \ref{figintegral} that the maximum value for the integral will be obtained at $m=0$, as it can be seen in the graduated axis that corresponds to $\sigma$. So we should take
\begin{align}
\sigma(\gamma^{2},0,0) & =2(2\pi)^{3}g^{2}N\int\frac{d^{3}k}{(2\pi)^{3}}\left(\frac{1}{k^{4}+\gamma^{4}}\right).
\end{align}
And so we are able to define the one loop gap equation :
\begin{equation}
2(2\pi)^{3}g^{2}N\int\frac{d^{3}k}{(2\pi)^{3}}\left(\frac{1}{k^{4}+\gamma^{4}}\right)=1.\label{eq:oneloopgapequation}
\end{equation}

Thus the $\gamma$ parameter is not independent, being defined as a function of the coupling
constant g:
\begin{equation}
\gamma=\sqrt{2}\pi^{2}Ng^{2}.
\end{equation}
It is clear that in close analogy to the Gribov-Zwanziger procedure \cite{Zwanziger2,Zwanziger25,Zwanziger3}
it is also desired to work directly with the gap equation $\frac{\delta\Gamma}{\delta\gamma^{2}}=0$. It is well known that due to the fact that $\Gamma$ corresponds to the energy in the euclidean case, $\varepsilon_{v}=0$. It is necessary to break the supersymmetry explicitly in order to use this gap equation. The most simple way in order to do that is introducing a supersymmetry breaking term only which acts changing the value of the Gribov parameter in the fermionic partners
\begin{equation}
S_{break}=\rho\int d^{3}x\overline{\lambda}^{a\alpha}\frac{\epsilon_{\alpha\beta}}{\partial^{2}}\lambda^{a\beta}.
\end{equation}
This term breaks the supersymmetry and introduces, at principle, two different mass parameters that can be associated to a Gribov type behaviour. The original $\gamma$ that appears in the gauge sector and the parameter $\rho$ in the fermionic sector. Now, the equation $\frac{\delta\Gamma}{\delta\gamma^{2}}=0$ is not trivial due to the fact that $\Gamma$ is not zero anymore but a function of $\gamma$ and $\rho$, $\Gamma(m,\rho,\gamma)$. Now the derivation in $\gamma$ corresponds to a filtration only in the gauge sector, i.e. in the sector in which the $\gamma$ appear. It is important to note that due to the structure of the Gribov propagator we have a gluino condensate given by $\lim_{x_{1}\rightarrow x_{2}}Tr <\lambda_{\alpha}^{a}(x_{1})\lambda_{\beta}^{b}(x_{2})>$, which corresponds to the term added to the fermionic action. Thus the use of the gap equation permits to obtain the same integral condition as explained in the  analysis of the ghost sector and used to obtain the value of $\gamma$. Again it is important to note that this procedure is equivalent to a filtration on gauge-ghost sector. At the end in order to recover the supersymmetry we impose $\Gamma(m,\rho(\gamma),\gamma)=0$ and obtain again the original propagator for the gluino sector, fixing the $\rho$ as the original value for this parameter that is obtained in the direct superfield formalism used in the construction of the Gribov term in section 3. Of course this method is much more adequate in the case of non explicit supersymmetric construction of the Gribov-Zwanziger procedure which is not the case in this work. The only advantage of the method of breaking and restoring the supersymmetry is to note that only the sector of the energy functional that is derived from the gauge fields contributes in order to obtain the value of $\gamma$.
The results will be the same as in the more simple method explained
in these section. The behaviour of the gauge and gaugino correlators is the same as presented in \cite{Capri:2014xea}. Of course due to the construction directly in superfields of our Gribov action it is not necessary the introduction, by hand, of the Gribov term for the fermionic sector, like in \cite{Capri:2014xea}. The price to pay for the explicitly supersymmetric construction is the need to look only to the ghost sector in order to fix the value of $\gamma$. In other terms, the explicitly superfield construction of the Gribov action gave us the prove that only the gauge-ghost sector is fundamental for the Gribov mechanism even in a supersymmetric action.

\section{Some aspects of N= 1 Chern-Simons-Matter Theories}

Now we will consider only the Chern-Simons sector. With interest in
BLG and ABJM theories. Many works about BLG and ABJM exists in the literature like \cite{Bagger, Maldacena, Gustavsson:2007vu} and recently about BRST breaking in ABJM theory \cite{Faizal}. So we would focus on the possibility of a replica model \cite{replica} for confinement using the two Chern-Simons sectors of theories of that type. So considering only the Chern-Simons sector, we have that the ultraviolet
mass dimension of gauge superfield $\Gamma_{\alpha}^{a}$ becomes
$\frac{1}{2}$. In this case we rewrite the action (\ref{eq:actionSCS})
as
\begin{equation}
S_{SCS}=ik\int d^{3}xd^{2}\theta\left[(D^{\alpha}\Gamma^{a\beta})(D_{\beta}\Gamma_{\alpha}^{a})+\frac{2}{3}if^{abc}\Gamma^{a\alpha}\Gamma^{b\beta}(D_{\beta}\Gamma_{\alpha}^{c})-\frac{1}{6}f^{abc}f^{cde}\Gamma^{a\alpha}\Gamma^{b\beta}\Gamma_{\alpha}^{d}\Gamma_{\beta}^{e}\right].\label{eq:actionSCS-1abjm}
\end{equation}
As we are interested in N = 1 supersymmetric gauge field theory with
the gauge group $G\times G$, we write a second action for another
gauge superfield $\tilde{\Gamma}_{\alpha}^{a}$ 
\begin{equation}
\tilde{S}_{SCS}=i\tilde{k}\int d^{3}xd^{2}\theta\left[(D^{\alpha}\tilde{\Gamma}^{a\beta})(D_{\beta}\tilde{\Gamma}_{\alpha}^{a})+\frac{2}{3}if^{abc}\tilde{\Gamma}^{a\alpha}\tilde{\Gamma}^{b\beta}(D_{\beta}\tilde{\Gamma}_{\alpha}^{c})-\frac{1}{6}f^{abc}f^{cde}\tilde{\Gamma}^{a\alpha}\tilde{\Gamma}^{b\beta}\tilde{\Gamma}_{\alpha}^{d}\tilde{\Gamma}_{\beta}^{e}\right].\label{eq:actionSCS-2abjm}
\end{equation}
And we take the following total Chern-Simons action with matter:
\begin{equation}
S=S_{SCS}-\tilde{S}_{SCS}+S_{matter}.\label{eq:actionABJM}
\end{equation}
With the matter action given by
\begin{equation}
S_{matter}=\int d^{3}xd^{2}\theta tr\left(\nabla^{\alpha}X^{I\dagger}\nabla_{\alpha}X_{I}+V\right),\label{eq:abjmmatter}
\end{equation}
with the matter superfield $X$ in the bi-fundamental representation
of the gauge group, i.e the superspace covariant derivatives for matrix-valued
complex scalar superfields $X^{I}$ and $X^{I\dagger}$ are defined
by 
\begin{eqnarray}
\nabla_{\alpha}X^{I} & = & D_{\alpha}X^{I}+i\Gamma_{\alpha}X^{I}-iX^{I}\tilde{\Gamma}_{\alpha},\nonumber \\
\nabla_{\alpha}X^{I\dagger} & = & D_{\alpha}X^{I\dagger}-iX^{I\dagger}\Gamma_{\alpha}+i\tilde{\Gamma}_{\alpha}X^{I\dagger},
\end{eqnarray}
and $V$ is the potential term given by 
\begin{eqnarray}
V & = & \frac{1}{k}\epsilon^{IJ}\epsilon_{KL}[X_{I}X^{K\dagger}X_{J}X^{L\dagger}].
\end{eqnarray}
The classical Action (\ref{eq:actionABJM}) remains invariant under
the following gauge transformation 
\begin{eqnarray}
\delta\Gamma_{\alpha}=\nabla_{\alpha}\Lambda, &  & \delta\tilde{\Gamma}_{\alpha}=\tilde{\nabla}_{\alpha}\tilde{\Lambda},\nonumber \\
\delta X^{I}=i(\Lambda X^{I}-X^{I}\tilde{\Lambda}), &  & \delta X^{I\dagger}=i(\tilde{\Lambda}X^{I\dagger}-X^{I\dagger}\Lambda),
\end{eqnarray}
where $\Lambda=\Lambda^{A}T_{A}$ and $\tilde{\Lambda}=\tilde{\Lambda}^{A}\tilde{T}_{A}$
are parameters of transformations. 
The gauge invariance of this theory reflects that the theory have
some spurious degrees of freedom. In order to quantize the theory
correctly we need to fix the gauge. And we can do with the Faddeev-Popov
method already studied, ending with an action of gauge fixing for
each superfield $\Gamma_{\alpha}^{a}$ and $\tilde{\Gamma}_{\alpha}^{a}$,
equation (\ref{eq:actiongaugefix}) \cite{Faizal}.

In terms of the field components this action can represents the gauge
part of the ABJM or BLG : with gauge group $G=SU(2)$, we can have
a decomposition of BLG theory (It is possible to decompose the gauge
symmetry generated by $SO(4)$ into $SU(2)\times SU(2)$) and with
$G=U(N)$, we can have ABJM theory \cite{Bagger,Maldacena,Gustavsson:2007vu,Faizal}.
In both cases we have, with $\tilde{k}=\pm k$, $G(N)_{k}\times G(N)_{\pm k}$,
with $k$ the Chern-Simons level.

We know that the solution of the Gribov problem for a Chern-Simons
theory is trivial because the theory does not have metric \cite{SorellaYMCSGribov}.
But now, in this case with two Chern-Simons interacting with bi-fundamental
matter, we can have a spontaneous symmetry breaking such that insert
a metric \cite{Mukhi2008, Ketov2011}. This open a possibility of seeing 
Gribov\footnote{Gribov confinement scenario at level of propagator, namely, 
if we can get a Gribov type propagator.} 
as a phase in ABJM theory, what we started to investigate here.

Here we only consider the two Chern-Simons actions in a toy model.
Consider then the following mixing terms between the gauge fields.
\begin{equation}
S_{mix}=\mu\int d^{3}xd^{2}\theta\left(\Gamma^{a\alpha}\Gamma_{\alpha}^{a}-\tilde{\Gamma}^{a\alpha}\tilde{\Gamma}_{\alpha}^{a}+2\Gamma^{a\alpha}\tilde{\Gamma}_{\alpha}^{a}\right).\label{eq:abjmmatter-1-1}
\end{equation}
Where $\mu$ has the dimension of mass. We expect it is possible
to get this mass and mixing term through spontaneous symmetry breaking \cite{AmaralLemes2015progress}. And first we will
address the following total action (levels $k$ and $\tilde{k}=-k$)
\begin{equation}
S=S_{SCS_{(k)}}-\tilde{S}_{SCS_{(k)}}+S_{mix}.\label{eq:actionABJM-1-1}
\end{equation}
So we get for full bilinear part of this total action 
\begin{align}
 & \int d^{3}xd^{2}\theta\left[ik(D^{\alpha}\Gamma^{a\beta})(D_{\beta}\Gamma_{\alpha}^{a})-ik(D^{\alpha}\tilde{\Gamma}^{a\beta})(D_{\beta}\tilde{\Gamma}_{\alpha}^{a})+\mu\left(\Gamma^{a\alpha}\Gamma_{\alpha}^{a}-\tilde{\Gamma}^{a\alpha}\tilde{\Gamma}_{\alpha}^{a}+2\Gamma^{a\alpha}\tilde{\Gamma}_{\alpha}^{a}\right)\right]\nonumber \\
= & \int d^{3}xd^{2}\theta\left[ik(\Gamma_{\beta}^{a}D^{\alpha}D^{\beta}\Gamma_{\alpha}^{a})-ik(\tilde{\Gamma}_{\beta}^{a}D^{\alpha}D^{\beta}\tilde{\Gamma}_{\alpha}^{a})+\mu\left(\varepsilon^{\alpha\beta}\Gamma_{\beta}^{a}\Gamma_{\alpha}^{a}-\varepsilon^{\alpha\beta}\tilde{\Gamma}_{\beta}^{a}\tilde{\Gamma}_{\alpha}^{a}+2\varepsilon^{\alpha\beta}\Gamma_{\beta}^{a}\tilde{\Gamma}_{\alpha}^{a}\right)\right].
\end{align}
And we get the propagators in momentum space of the type of (\ref{eq:SYMCSpropagatorGZ})
\begin{eqnarray}
<\Gamma_{\alpha}^{a}(1)\Gamma_{\beta}^{b}(2)> & = & \frac{1}{8k^{2}}\delta^{ab}\left[\frac{(2kp^{2}+\frac{2}{k}\mu^{2})iD^{2}+(p^{2}+\frac{1}{2k^{2}}\mu^{2})\mu}{(p^{4}+\frac{1}{4k^{4}}\mu^{4})}\right]\frac{D^{2}}{p^{2}}D_{\beta}D_{\alpha}\delta^{2}(\theta_{1}-\theta_{2})\delta^{3}(x_{1}-x_{2}),\nonumber \\
<\Gamma_{\alpha}^{i}(1)\Gamma_{\beta}^{j}(2)> & = & <\tilde{\Gamma}_{\alpha}^{i}(1)\tilde{\Gamma}_{\beta}^{j}(2)>,\label{eq:ABJMpropagator}
\end{eqnarray}
and a mixing propagator
\begin{eqnarray}
<\Gamma_{\alpha}^{a}(1)\tilde{\Gamma}_{\beta}^{b}(2)> & = & -\frac{\mu}{8k^{2}}\delta^{ab}\left[\frac{p^{2}-\frac{1}{2k^{2}}\mu^{2}-\frac{1}{k}i\mu D^{2}}{(p^{4}+\frac{1}{4k^{4}}\mu^{4})}\right]\frac{D^{2}}{p^{2}}D_{\beta}D_{\alpha}\delta^{2}(\theta_{1}-\theta_{2})\delta^{3}(x_{1}-x_{2}).
\end{eqnarray}
In the strong coupling regime (taking $k=1$) this propagators have
the pole of Gribov type.

Let's address now the action in a regime with levels $k$ and $\tilde{k}=k$:
\begin{equation}
S=S_{SCS_{(k)}}+\tilde{S}_{SCS_{(k)}}+S_{mix}.\label{eq:actionABJM-1-1-1}
\end{equation}
We get for full bilinear part of this total action 
\begin{align}
 & \int d^{3}xd^{2}\theta\left[ik(D^{\alpha}\Gamma^{a\beta})(D_{\beta}\Gamma_{\alpha}^{a})+ik(D^{\alpha}\tilde{\Gamma}^{a\beta})(D_{\beta}\tilde{\Gamma}_{\alpha}^{a})+\mu\left(\Gamma^{a\alpha}\Gamma_{\alpha}^{a}-\tilde{\Gamma}^{a\alpha}\tilde{\Gamma}_{\alpha}^{a}+2\Gamma^{a\alpha}\tilde{\Gamma}_{\alpha}^{a}\right)\right]\nonumber \\
= & \int d^{3}xd^{2}\theta\left[ik(\Gamma_{\beta}^{a}D^{\alpha}D^{\beta}\Gamma_{\alpha}^{a})+ik(\tilde{\Gamma}_{\beta}^{a}D^{\alpha}D^{\beta}\tilde{\Gamma}_{\alpha}^{a})+\mu\left(\varepsilon^{\alpha\beta}\Gamma_{\beta}^{a}\Gamma_{\alpha}^{a}-\varepsilon^{\alpha\beta}\tilde{\Gamma}_{\beta}^{a}\tilde{\Gamma}_{\alpha}^{a}+2\varepsilon^{\alpha\beta}\Gamma_{\beta}^{a}\tilde{\Gamma}_{\alpha}^{a}\right)\right].
\end{align}
And we get the propagators in momentum space:
\begin{eqnarray}
<\Gamma_{\alpha}^{a}(1)\Gamma_{\beta}^{b}(2)> & = & \frac{1}{8k^{2}}\delta^{ab}\left[\frac{(2kp^{2}-\frac{1}{k}\mu^{2})iD^{2}+(p^{2}+\frac{1}{2k^{2}}\mu^{2})\mu}{(p^{2}+\frac{1}{2 k^{2}}\mu^{2})^{2}}\right]\frac{D^{2}}{p^{2}}D_{\beta}D_{\alpha}\delta^{2}(\theta_{1}-\theta_{2})\delta^{3}(x_{1}-x_{2}),\nonumber \\
<\Gamma_{\alpha}^{i}(1)\Gamma_{\beta}^{j}(2)> & = & <\tilde{\Gamma}_{\alpha}^{i}(1)\tilde{\Gamma}_{\beta}^{j}(2)>,\label{eq:ABJMpropagator-1}
\end{eqnarray}
and a mixing propagator
\begin{eqnarray}
<\Gamma_{\alpha}^{a}(1)\tilde{\Gamma}_{\beta}^{b}(2)> & = & -\frac{\mu}{8k^{2}}\delta^{ab}\left[\frac{1}{p^{2}+\frac{1}{2k^{2}}\mu^{2}}\right]\frac{D^{2}}{p^{2}}D_{\beta}D_{\alpha}\delta^{2}(\theta_{1}-\theta_{2})\delta^{3}(x_{1}-x_{2}).
\end{eqnarray}
This regime is responsible for massive particles. So we obtain in this toy model at least two phases, one of which is confinement in Gribov-Zwanziger scenario.

%The meaning of this case have only a term of type SYM and still need $k$ imaginary,  demands further research for better understanding.
%The possibility of obtaining a dynamical mechanism in order to turn this phase transition into a dynamical one is really a difficult task and certainly is a point to be studied in a future work.

\section{Conclusions}

In this work we have studied the Super-Yang-Mills Chern-Simons theory,
N = 1, considering the Gribov problem present in YM theories as well
as a generalization of the Gribov-Zwanziger approach with auxiliary
superfields in superspace. 
A local supersymmetric Gribov-Zwanziger sector is presented providing the starting point in order 
to implement the restriction to the first Gribov region beyond one-loop order. 
Which results in propagators of the Gribov type opening a perspective of treating confinement in these theories
 in terms of  Gribov-Zwanziger scenario. And we note that Gribov acts 
 as a regulator for the usual infrared SYM-CS theory, which can be observed from 
 super propagator (\ref{eq:SYMCSpropagatorGZ})
Also a possible mechanism in order to obtain a Gribov like propagator from a ABJM theory is presented.
It suggests that it is possible by making use of a 
spontaneous symmetry breaking mechanism similar to the one 
proposed in \cite{Amaral:2013uja}, which is under investigation. 
 This mechanism can offer a possibility of seeing Gribov as a phase in ABJM theory, 
 offering a link between confinement in Gribov-Zwanziger scenario and confinement 
 as seeing by the brane scenario. We expect to study the implications of this relation 
 in future works, specially the possibility of obtaining a condensate that can be associated 
 to a usual particle in K\"{a}llen-Lehmann representation.

\section*{Acknowledgements}

The Conselho Nacional de Desenvolvimento Cientfico e tecnolgico
CNPq- Brazil, Fundao de Amparo a Pesquisa do Estado do Rio de Janeiro
(Faperj). the SR2-UERJ and the Coordenao de Aperfeioamento de Pessoal
de Nvel Superior (CAPES) are acknowledged for the financial support. 

\appendix
%dummy comment inserted by tex2lyx to ensure that this paragraph is not empty

\section[Appendix]{Notation, conventions and some useful formulas}

We work with Euclidean metric: diag(+++). So we choose the gamma matrices
being the Pauli matrices $\sigma_{i}$ {[}\cite{GradedMajorana}{]}:

\begin{equation}
\gamma^{\mu}\equiv(\sigma_{\mu})_{\alpha}^{\:\beta}
\end{equation}
witch are OS self-conjugate and:

\begin{equation}
\{\sigma^{\mu},\sigma^{\nu}\}=2\delta^{\mu\nu}I,
\end{equation}
\begin{equation}
[\sigma^{\mu},\sigma^{\nu}]=2i\varepsilon^{\mu\nu\sigma}\sigma^{\sigma}.
\end{equation}
The invariant anti-symmetric tensor is defined as
\begin{equation}
\varepsilon^{-+}=\varepsilon_{-+}=+1,
\end{equation}
\begin{equation}
\varepsilon^{\gamma\beta}\varepsilon_{\beta\alpha}=-\delta_{\alpha}^{\gamma},\label{eq:epsoncontraction}
\end{equation}
and are used to raise and lower indices as conversion:
\begin{equation}
\psi^{\alpha}=\varepsilon^{\alpha\beta}\psi_{\beta},
\end{equation}
\begin{equation}
\psi_{\alpha}=\psi^{\beta}\varepsilon_{\beta\alpha}.
\end{equation}

In this way is possible to find the representation of differential
operator of the generators of super algebra in D=3, with the concept
of graded Majorana \cite{GradedMajorana}:
\begin{equation}
Q_{\alpha}=-\partial_{\alpha}+\partial_{\alpha\beta}\theta^{\beta},
\end{equation}
with
\begin{equation}
\partial_{\alpha\beta}=i\sigma_{\alpha}^{\mu\gamma}\varepsilon_{\gamma\beta}\partial_{\mu}.
\end{equation}
As well as the superspace derived:
\begin{equation}
D_{\alpha}=\partial_{\alpha}+\partial_{\alpha\beta}\theta^{\beta},
\end{equation}
with the following relations:
\begin{equation}
\{D_{\alpha},D_{\beta}\}=2\partial_{\alpha\beta},
\end{equation}
\begin{equation}
[D_{\alpha},D_{\beta}]=-2\varepsilon_{\alpha\beta}D^{2},
\end{equation}
\begin{equation}
D_{\alpha}D_{\beta}=\partial_{\alpha\beta}-\varepsilon_{\alpha\beta}D^{2},\label{eq:superderivativealfabetarelation}
\end{equation}
\begin{equation}
D^{\beta}D_{\alpha}D_{\beta}=0.\label{eq:superderivativealfabetaalfarelationzero}
\end{equation}
And it is easy to verify that
\begin{equation}
[Q_{\alpha},D_{\beta}]=0.
\end{equation}

Another useful relations:

\begin{equation}
\partial_{\alpha\beta}\partial^{\alpha\gamma}=\partial^{2}\delta_{\beta}^{\gamma},\label{eq:realtionderivativecontraction}
\end{equation}
\begin{equation}
(D^{2})^{2}=-\partial^{2},\label{eq:D2squared}
\end{equation}
\begin{equation}
\int d^{2}\theta=-\frac{1}{4}D^{2},
\end{equation}

\end{document}